\documentclass{iau}

\usepackage{amsmath}
\usepackage{graphicx}
\usepackage{multirow}
\usepackage{natbib}

\begin{document}

\lefttitle{Sergio Mart\'in et al.}
\righttitle{Chemistry here and there: Comparison between the CMZ of the Milky Way and NGC 253}

\jnlPage{1}{7}
\jnlDoiYr{2021}
\doival{10.1017/xxxxx}

\aopheadtitle{Proceedings IAU Symposium}
\editors{M. Zaja\v{c}ek,  T. Je\v{r}\'{a}bkov\'{a}, V. Karas, R. Schödel \&  P. Sukov\'{a}, eds.}

\title{Chemistry here and there: Comparison between the CMZ of the Milky Way and NGC 253}

%\author{Sergio Mart\'in $^1,^2$, Alvaro L\'opez-Gallifa $^3$, David San Andr\'es $^3$, Victor Rivilla $^3$, and ALCHEMI collaboration}
%\affiliation{$^1$ European Southern Observatory, Alonso de Córdova, 3107, Vitacura, Santiago 763-0355, Chile}
%\affiliation{$^2$ Joint ALMA Observatory, Alonso de Córdova, 3107, Vitacura, Santiago 763-0355, Chile}
%\affiliation{$^3$Centro de Astrobiología (CAB) CSIC-INTA, Ctra. de Ajalvir, km. 4, Torrejón de Ardoz, E-28850 Madrid, Spain}

\author{Sergio Mart\'in $^1,^2$, Alvaro L\'opez-Gallifa $^3$, David San Andr\'es $^3$, Victor Rivilla $^3$, and ALCHEMI collaboration}
\affiliation{$^1$ European Southern Observatory, Alonso de Córdova, 3107, Vitacura, Santiago 763-0355, Chile\\[3pt]
	$^2$ Joint ALMA Observatory, Alonso de Córdova, 3107, Vitacura, Santiago 763-0355, Chile\\[3pt]
	$^3$Centro de Astrobiología (CAB) CSIC-INTA, Ctra. de Ajalvir, km. 4, Torrejón de Ardoz, E-28850 Madrid, Spain}

\begin{abstract}
For decades, chemical studies in external galaxies have been limited by sensitivity, resolution, bandwidth or a combination of them. A decade of ALMA operations, made the detailed study of the molecular abundances in nearby galaxies a routine job. We can finally bridge the ISM chemistry from Galactic to extragalactic environments, through one to one comparison, albeit probing significantly different scales.
The ALMA large program ALCHEMI imaged the chemistry of the starburst galaxy NGC 253 at an unprecedented combination of sensitivity and resolution, achieving a detailed spatially resolved chemical inventory. ALCHEMI probed molecular differences among regions within its CMZ, and allowed the comparison with Galactic environments ranging from hot cores, quiescent molecular clouds and comets. This comparison can be put in the context of available data towards other extragalactic environments.
Globally, the comparisons presented here show an astoundingly good correlation between the chemistry in Galactic GMCs and the central molecular zone of NGC~253, ranging scales from 1~pc up to $\sim 0.5$~kpc.
The wideband sensitivity upgrade of ALMA will expand similar comparisons towards a wider range of Galactic and extragalactic objects.
\end{abstract}


\maketitle

\section{Towards a chemical classification of galaxies}

Over the past two decades, systematic deep molecular studies have been performed towards the brightest nearby extragalactic emitters \citep[see Table 1 in][]{Martin2021}. Some of these works aimed to achieve a chemical classification of galaxies, attending to their nuclear power source \citep[e.g.][]{Mart'in2006,Aladro2011,Takano2019}. Although the effect of the pervading UV field has a clear imprint in the molecular abundances along the starburst evolutionary path, this differentiation is not so obvious when compared with AGN dominated galaxies or strongly obscured ULIRGs \citep{Martin2009,Mart'in2011, Aladro2015, Costagliola2015}. Moreover, a significant resemblance was reported with the quiescent molecular gas within the Galaxy \citep{Mart'in2006, Muller2011}, where we can potentially explore what is behind this apparent homogeneous chemistry observed in the extragalactic medium.

The advent of high sensitivity broadband interferometers like ALMA and NOEMA has allowed the routine observation the central molecular zone of nearby galaxies at a few tens of parsec spatial resolution \citep[e.g.][]{Harada2019,Stuber2025,Eibensteiner2022} as well as pushing the limits of chemical complexity to high redshift galaxies \citep[e.g.][]{Yang2023}.

\section{The ALCHEMI project}

The ALMA Comprehensive High-resolution Extragalactic Molecular Inventory \citep[ALCHEMI;][]{Martin2021} project, was an ALMA large program designed not only to have the most comprehensive molecular study at giant molecular cloud (GMC) scale in an extragalactic starbursting environment, but to enable an astrochemical connection to the Galactic center (GC). The target was the nearby prototypical starburst galaxy NGC~253, one of the most prolific extragalactic molecular emitters, and consisted of an unbiased spectral line imaging survey (84 to 374~GHz, ranging ALMA Bands 3 through 7) of the whole central molecular zone (CMZ, $\rm 600~pc \times 300~pc$) at an unprecedented combination of resolution ($1.6^{\prime\prime}$) and sensitivity ($10-20$~mK). The ALCHEMI collaboration has produced more than 20 publications \footnote{\url{https://scixplorer.org/public-libraries/gPMmTM9DSaSyNk7yx6HCsg}}, spanning from continuum multi-wavelength analysis \citep{Humire2025} to the use of machine learning algorithms for spectral and chemical model fitting \citep{Barrientos2021,Behrens2024}. 

Here we briefly summarize some results relevant for the comparison between our GC and the extragalactic environment.

Of particular relevance is the detection of Phosphorus Nitride (PN) due to its relevance as prebiotic chemistry precursor. The observed subthermal excitation and PN/SiO abundance ratio both follow the trend found towards the Galactic Center molecular clouds, confirming the shock origin of PN in both environments \citep{Haasler2022}.

In fact, the shocks across the whole CMZ of NGC~253 probed through various molecular proxies \citep{Harada2022,Gong2025}, and its evolution was studied through the SLED analysis of HNCO and SiO, under the assumption of both species being originated in a single shock event \citep{Huang2023}. While SiO traces fast shocks with kinetic temperatures of a few 100~K, HNCO traces denser and cooler slow shocks with $T_{kin}\sim 100$~K.

Whithin the Galactic center CMZ, the comparison of HNCO emission and the magnetic field orientation shows how the changing magnetic field versus HNCO filaments orientations affects turbulence, which can assist or inhibit cloud collapse \citep[see Posters S30 and S31 by Dylan Par\'e,][]{Pare2026}. Although the resolution of 0.05~pc towards the GC is still out of reach even in nearby active galaxies, the ALCHEMI resolution or $\sim28~pc$ approaches past GC studies with single dish observations \citep[$\sim1.5$~pc,][]{Jones2012}, and could potentially be combined with magnetic field measurements \citep{Belfiori2025} to evaluate average properties towards NGC~253.

Various ALCHEMI works made use of different molecular probes and their comparison to state-of-the-art chemical models to derive the pervading cosmic ray ionization rate. All these probes, C$_2$H abundances \citep{Holdship2021}, HCO$^+$/HOC$^+$ ratios \citep{Harada2021}, H$_3$O$^+$/SO ratio \citep{Holdship2022}, and HCN/HNC ratio \citep{Behrens2022,Behrens2024}, yield similarly extreme rates of $\zeta\sim10^{3-4}\zeta_0$, where $\zeta_0\sim10^{-17}\rm s^{-1}$ is the Milky Way canonical rate towards the Galactic spiral arms. More importantly, this is similar or even an order of magnitude higher than the rate towards the Galactic center \citep[Talk by Sruthiranjani Ravikularaman,][]{Oka2005,Oka2019}

However, the largest potential of unbiased spectral line surveys like ALCHEMI results from the study of multiple molecular probes. \citet{Tanaka2024} modeled 11 bright dense molecular tracers to image the physical properties across the CMZ and compared the properties with those found in the GC. NGC~253 showed ten times more high density gas and three times higher dense mass fraction. While this is not enough to justify the 30 times higher star forming rate in NGC~253, the larger fraction of gas with densities above $10^{4-5}\rm cm^{-3}$ might be the key for this star formation enhancement.

The work by \citet{Harada2024} embarked into the largest sample of molecular species and unblended transitions (44 and 148, respectively) ever used into a Principal Component Analysis (PCA) for any Galactic or extragalactic source. The results showed the clear differences between the species tracing young and evolved starburst regions, as well as the different morphologies of the low velocity shocks and star formation dominated locations. More importantly, it could be appreciated the evolution of the projection of the individual transitions of a given species into the different principal components as a function of its upper energy level. The follow up work by \citep{Kishikawa2025} showed the potential of the non-negative matrix factorization (NMF) in better disentangling the different structure components, usually appearing combined as positive or negative structures in the principal components in PCA.

\section{ALCHEMI vs GC: The e-galactic bridge}

\begin{figure}[t]
	\centerline{\vbox to 3pc{\hbox to 10pc{}}}
	\includegraphics[width=\textwidth]{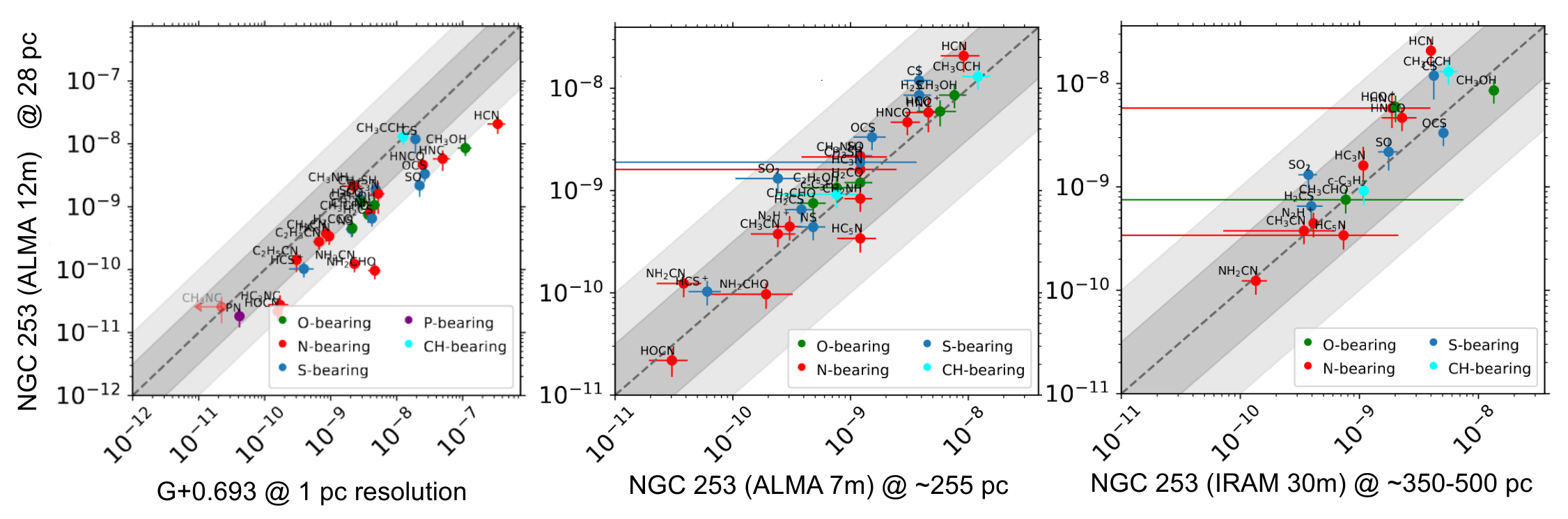}
	\caption{Comparision of molecular abundanced towards NGC~253 central molecular zone at varios spatial scales, from 28~pc to $\sim 500$~pc, as well as the comparision with the Galactic center giant molecular cloud G+0.693-0.027 (left panel). Dark and light grey zones indicate the differences of a factor of 3 and 10, respectively, from the 1-to-1 relation, shown as a dashed line. These comparison plots show the astoundingly good correlation between the Galactic GMC and the central molecular zone of NGC~253, ranging scales from 1~pc up to $\sim 0.5$~kpc. Figure adapted from L\'opez-Gallifa et al. (in Prep.)}
	\label{NGC253 comparison}
\end{figure}

ALCHEMI provides us with the unique opportunity to perform a one-to-one comparison with similar broadband millimeter spectral surveys existing towards Galactic sources. The work by L\'opez-Gallifa et al. (Poster S27, in Prep.) aimed to perform such a comparison using the unique molecular sample at hand towards NGC~253, finally making the bridge between Galactic and extra-galactic astrochemisty. This work, presented here, aimed to obtain an statistical comparison of the global molecular footprint in an starbursting environment, similar to that performed within the Galaxy \citep{LopezGallifa2024,LopezGallifa2025}, and did not aim to study in detail the deviations from any given molecular species.

The study by L\'opez-Gallifa et al. (Poster S27, in Prep.) analyzed spectra towards the 4 brighter GMCs in the CMZ of NGC~253 to analyze their molecular composition. Although the model included more than 150 species, only 35 of them were used in the comparison with Galactic sources. Abundances were opacity corrected and when available, the optically thinner isotopologues were used to derive the column densities of the brightest, optically thick, species. Also isotopic ratios were calculated for all observed isotopologues, providing accurate estimates of atomic isotopic ratios.

\begin{figure}[t]
	\centerline{\vbox to 3pc{\hbox to 10pc{}}}
	\includegraphics[width=\textwidth]{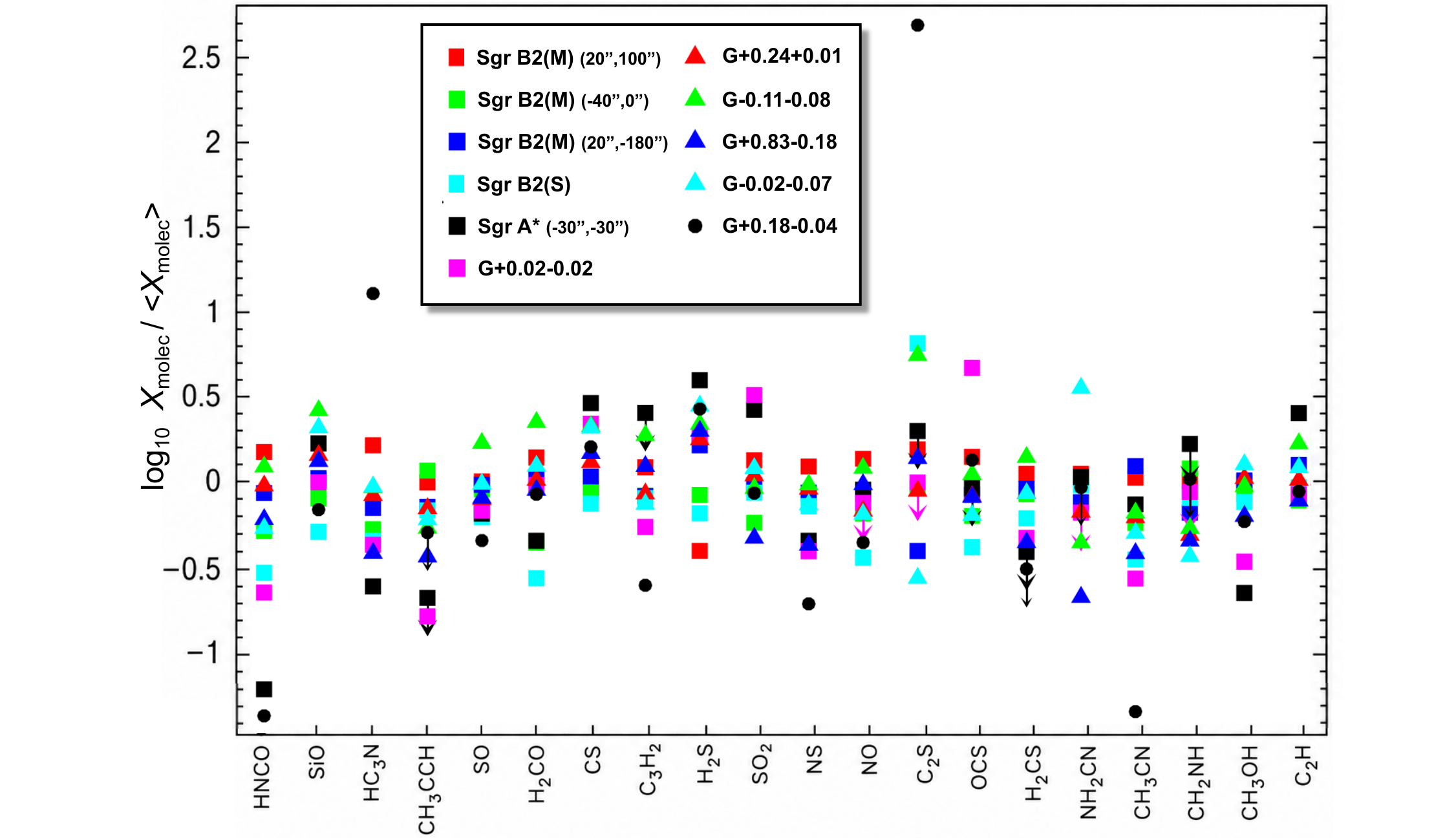}
	\caption{Molecular abundances of 20 species in a sample of 11 Galactic center molecular clouds. The abundances are referred to the average abundance of all detected molecules. The naming of sources follow \citet{Martin2008}, where a description of each source can also be found. Numbers in brackets correspond to offset from the named source in arcseconds. This comparison shows the relatively homogeneous chemistry within an order of magnitude across the Galactic center, with HNCO being the only molecule showing the clearer contrast towards UV radiated sources (black square and dot symbols).
	This Figure has been adapted from \citet{MartinRuiz2006}.
	}
	\label{GCcomparison}
\end{figure}

The first result of this work is the astounding homogeneity observed across the CMZ. All analyzed GMCs show a very close correlation of their molecular abundances indicating a homogeneous chemistry. This similarity also holds at different scales when compared the 28~pc resolution data from ALCHEMI, with the 7~m data alone at $\sim255$~pc resolution, or IRAM~30m single dish observations at $\sim350-500$~pc scales as shown in the right panels in Fig.~\ref{NGC253 comparison}.
This result is not different from what had been previously reported towards the center of the Milky-Way. The study from \citet{Requena-Torres2006} using a sample of 8 organic species towards 40 GC molecular clouds showed uniform abundance ratios both in GC clouds and Galactic hot cores. Moreover it was claimed that this might imply a similar average composition of grain mantles in both types of regions. Similarly, a study using 20 molecular species over a more limited sample of 11 GC clouds further supported this homogeneity \citep[Fig.~\ref{GCcomparison},][]{MartinRuiz2006}. That study showed that HNCO, among all sample species, showed the largest abundance dynamic range, turning this species into one of the best diagnostic probe to trace dense shocked gas with high degree of shielding against the pervading UV radiation from newly formed stars. In fact, the results displayed in Fig.~\ref{GCcomparison} were the origin of subsequent Galactic \citep{Martin2008} and extragalactic environments \citep{Martin2009}.

Even more interesting that the relative homogeneity across the central molecular zone of NGC~253, is the comparison with the Galactic center GMC G+0.693-0.027 (hereafter G+0.693). This GC source is located in the Sgr~B2 complex, north of the B2(N) and B2(M) hot cores. While there is not trace of star formation in this cloud, its chemistry appears to be driven by the low-velocity shocks originated by large-scale cloud-cloud collisions \citep{Zeng2020}. This source has become one of the most promising laboratories for the search of complex organic molecules in the interstellar medium (ISM, Poster S12 by Victor Rivilla). In fact, more than 30 species have been detected over the past few years towards G+0.693 \citep{Araki2026, SanzNovo2026}.

\begin{figure}[t]
	\centerline{\vbox to 3pc{\hbox to 10pc{}}}
	\includegraphics[width=\textwidth]{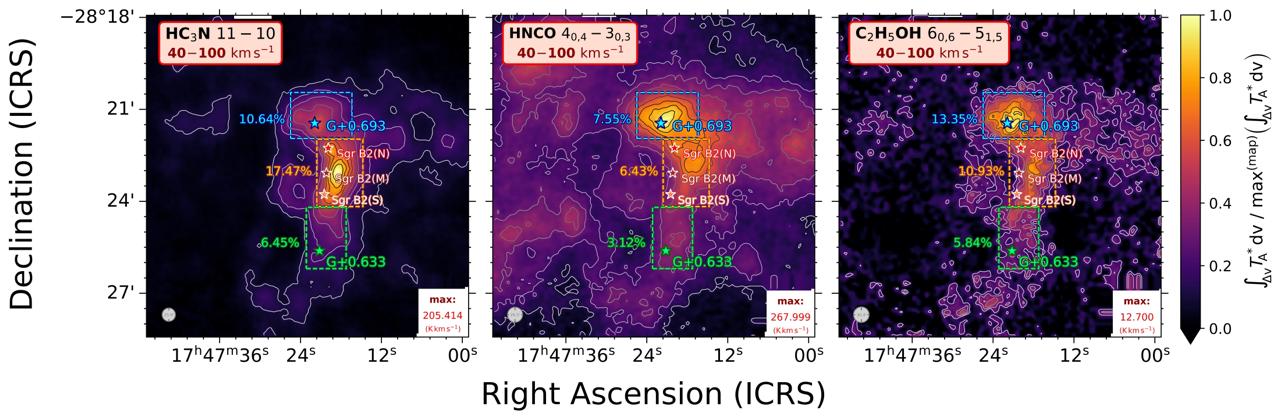}
	\caption{IRAM 30~m mosaic observations of the Sgr~B2 molecular complex showing the integrated emission of HC$_3$N, HNCO and C$_2$H$_5$OH. 
		The three dashed boxes show the contribution (in percentage) to the total mapped integrated emission from the regions around G+0.693 (blue), the hot cores (gold) and G+0.633 (green).
		%The three boxes show the percentage of the total integrated emission in the region around the hot cores, and the molecular clouds G+0.693 and G+0.633. 
		It is shown that most of the emission of these species is spread over large scales in this region. The Figure makes use of the data presented in \citet{SanAndres2026}.}
	\label{SgrB2Complex}
\end{figure}

The similarity between the observed chemistry in G+0.693 and NGC~253 over two orders of magnitude in spatial scales implies that the complex chemistry must be dominated by very extended emission and not constrained to the densest star forming cores even in the starburst environment.
In fact the single dish observation of the whole Sgr~B2 region \citep{SanAndres2026} show that the integrated molecular emission in this complex is mostly extended and only a relatively small fraction of the emission is concentrated in the hot cores and the molecular clouds G+0.693 and G+0.633 (the twin molecular cloud south of the Sgr~B2 hot cores, Talk by David San Andr\'es). As shown in Fig.~\ref{SgrB2Complex}, only $\sim35\%$ of the integrated emission of the dense gas tracer HC$_3$N is located within the hot cores and the two GMCs. More interestingly, an even lower percentage ($\sim17\%$) of the HNCO emission is observed not to be widespread, indicating the the vast majority of the chemical complexity in this region is released by shocks over extremely extended regions.

Such extended emission dominating the central molecular zone of NGC~253 is also found to be similar to that observed in other galactic nuclei, either dominated by star formation or by a nuclear active galactic nuclei, since the hot-core or AGN effect on the chemistry is restricted to much smaller scales. As such, the emission towards various types of nearby galactic nuclei appears to be dominated by G+0.693-like GMCs.

The work by L\'opez-Gallifa et al. (in Prep.) presented here and the previous Galactic studies mentioned above appear to suggest a homogeneous chemistry driven by a ``Universal'' dust chemistry, released into the ISM by large galactic scales shocks. This result does not imply that specific local physical conditions cannot be probed by the chemistry of specific species, since the deviation from the unity in Fig.~\ref{NGC253 comparison} can be of more than an order of magnitude. This result implies that globally, the observed chemistry is dominated by extended shock dominated molecular gas, and it is observed to be statistically similar across a wide range of scales. 

\section{A new beginning for astrochemistry at all scales}

The ALMA wideband sensitivity upgrade (WSU) will be yet another turning point in the field of astrochemistry. This upgrade will affect the whole ALMA signal chain from receiver upgrades, digital signal processing, transmission, down correlator and computing infrastructure.
This massive upgrade endeavor will provide an increase in the correlated bandwidth and sensitivity which will be crucial for the performance of ALMA in spectral scans over large bandwidths, needed for accurate astrochemical observations.

Such upgrade will enable ALCHEMI-like observations towards a large sample of galactic nuclei of different luminosities and activities, and even more, will enable large scale mapping of the Galactic central molecular zones like the ACES project (Talk by Steve Longmore) spanning over whole atmospheric windows or across the whole ALMA observable wavelengths.

%\begin{thebibliography}{}
%\bibitem[Bouvier(2013)]{2013EAS....62..143B} Bouvier, J.\ 2013, EAS Publications Series, 143
%\bibitem[Collier Cameron(1999)]{1999ASPC..158..146C} Collier Cameron, A.\ 1999, Solar and Stellar Activity: Similarities and Differences, 146
%\bibitem[Donati~{\it et. al}(1992)]{1992A&A...265..682D} Donati, J.-F., Brown, S.~F., Semel, M., {\it et. al}\ 1992, {\it A\&A}, 265, 682
%\end{thebibliography}
%
% Bibliography and bibfile
\def\aj{AJ}%
          % Astronomical Journal
\def\actaa{Acta Astron.}%
          % Acta Astronomica
\def\araa{ARA\&A}%
          % Annual Review of Astron and Astrophys
\def\apj{ApJ}%
          % Astrophysical Journal
\def\apjl{ApJ}%
          % Astrophysical Journal, Letters
\def\apjs{ApJS}%
          % Astrophysical Journal, Supplement
\def\ao{Appl.~Opt.}%
          % Applied Optics
\def\apss{Ap\&SS}%
          % Astrophysics and Space Science
\def\aap{A\&A}%
          % Astronomy and Astrophysics
\def\aapr{A\&A~Rev.}%
          % Astronomy and Astrophysics Reviews
\def\aaps{A\&AS}%
          % Astronomy and Astrophysics, Supplement
\def\azh{AZh}%
          % Astronomicheskii Zhurnal
\def\baas{BAAS}%
          % Bulletin of the AAS
\def\bac{Bull. astr. Inst. Czechosl.}%
          % Bulletin of the Astronomical Institutes of Czechoslovakia 
\def\caa{Chinese Astron. Astrophys.}%
          % Chinese Astronomy and Astrophysics
\def\cjaa{Chinese J. Astron. Astrophys.}%
          % Chinese Journal of Astronomy and Astrophysics
\def\icarus{Icarus}%
          % Icarus
\def\jcap{J. Cosmology Astropart. Phys.}%
          % Journal of Cosmology and Astroparticle Physics
\def\jrasc{JRASC}%
          % Journal of the RAS of Canada
\def\mnras{MNRAS}%
          % Monthly Notices of the RAS
\def\memras{MmRAS}%
          % Memoirs of the RAS
\def\na{New A}%
          % New Astronomy
\def\nar{New A Rev.}%
          % New Astronomy Review
\def\pasa{PASA}%
          % Publications of the Astron. Soc. of Australia
\def\pra{Phys.~Rev.~A}%
          % Physical Review A: General Physics
\def\prb{Phys.~Rev.~B}%
          % Physical Review B: Solid State
\def\prc{Phys.~Rev.~C}%
          % Physical Review C
\def\prd{Phys.~Rev.~D}%
          % Physical Review D
\def\pre{Phys.~Rev.~E}%
          % Physical Review E
\def\prl{Phys.~Rev.~Lett.}%
          % Physical Review Letters
\def\pasp{PASP}%
          % Publications of the ASP
\def\pasj{PASJ}%
          % Publications of the ASJ
\def\qjras{QJRAS}%
          % Quarterly Journal of the RAS
\def\rmxaa{Rev. Mexicana Astron. Astrofis.}%
          % Revista Mexicana de Astronomia y Astrofisica
\def\skytel{S\&T}%
          % Sky and Telescope
\def\solphys{Sol.~Phys.}%
          % Solar Physics
\def\sovast{Soviet~Ast.}%
          % Soviet Astronomy
\def\ssr{Space~Sci.~Rev.}%
          % Space Science Reviews
\def\zap{ZAp}%
          % Zeitschrift fuer Astrophysik
\def\nat{Nature}%
          % Nature
\def\iaucirc{IAU~Circ.}%
          % IAU Cirulars
\def\aplett{Astrophys.~Lett.}%
          % Astrophysics Letters
\def\apspr{Astrophys.~Space~Phys.~Res.}%
          % Astrophysics Space Physics Research
\def\bain{Bull.~Astron.~Inst.~Netherlands}%
          % Bulletin Astronomical Institute of the Netherlands
\def\fcp{Fund.~Cosmic~Phys.}%
          % Fundamental Cosmic Physics
\def\gca{Geochim.~Cosmochim.~Acta}%
          % Geochimica Cosmochimica Acta
\def\grl{Geophys.~Res.~Lett.}%
          % Geophysics Research Letters
\def\jcp{J.~Chem.~Phys.}%
          % Journal of Chemical Physics
\def\jgr{J.~Geophys.~Res.}%
          % Journal of Geophysics Research
\def\jqsrt{J.~Quant.~Spec.~Radiat.~Transf.}%
          % Journal of Quantitiative Spectroscopy and Radiative Trasfer
\def\memsai{Mem.~Soc.~Astron.~Italiana}%
          % Mem. Societa Astronomica Italiana
\def\nphysa{Nucl.~Phys.~A}%
          % Nuclear Physics A
\def\physrep{Phys.~Rep.}%
          % Physics Reports
\def\physscr{Phys.~Scr}%
          % Physica Scripta
\def\planss{Planet.~Space~Sci.}%
          % Planetary Space Science
\def\procspie{Proc.~SPIE}%
          % Proceedings of the SPIE

\bibliographystyle{aa}	% see astronat package, apj.bst
\bibliography{IAUS405_SergioMartin}

\end{document}